\documentclass[aps,prc,preprint,showpacs,preprintnumbers,amsmath,amssymb] {revtex4-1}
\usepackage{amsmath}
\usepackage{amssymb}
\usepackage{graphicx}
\usepackage{subfigure}
\usepackage{rotating}
\usepackage{natbib}
\usepackage{txfonts}
\usepackage{color}

\def\qbar{\bar{q}}
\def\Qbar{\overline{Q}}

\newcommand{\be}{\begin{equation}}
\newcommand{\ee}{\end{equation}}
\newcommand{\bge}{\begin{equation}}
\newcommand{\ene}{\end{equation}}
\newcommand{\bea}{\begin{eqnarray}}
\newcommand{\eea}{\end{eqnarray}}
\newcommand{\bg}{\begin{eqnarray}}
\newcommand{\en}{\end{eqnarray}}

\renewcommand{\vec}[1]{\boldsymbol{#1}}

\newcommand{\im}{\text{i}}

\def\eqref#1{Eq.~(\ref{eq:#1})}

\def\be{\begin{equation}}
\def\ee{\end{equation}}
\def\bg{\begin{eqnarray}}
\def\en{\end{eqnarray}}
\def\nn{\nonumber}

\long\def\Omit#1{}

\DeclareGraphicsExtensions{ .pdf, .jpg, .eps}
\begin{document}
\title{${\bar D}D$ meson pair production in antiproton-nucleus collisions 
}
\author{R.~Shyam$^{1}$}
\author{K.~Tsushima$^2$}
\affiliation{$^1$Saha Institute of Nuclear Physics, 1/AF Bidhan Nagar, Kolkata 700064, 
India}
\affiliation{$^2$Laboratorio de Fisica Teorica e Computacional,
Universidade Cruzeiro do Sul, Rua Galvao Bueno, 868, Liberdade
01506-000, Sao Paulo, SP, Brazil 
} 

\date{\today}
\begin{abstract}
We study the $\bar D D$ (${\bar D}^0 D^0$ and $D^-D^+$) charm meson pair 
production in antiproton (${\bar p}$) induced reactions on nuclei at 
beam energies ranging from threshold to several GeV. Our model is based on 
an effective Lagrangian approach that has only the baryon-meson degrees of 
freedom and involves the physical hadron masses. The reaction proceeds via 
the $t$-channel exchanges of $\Lambda_c^+$, $\Sigma_c^+$, and $\Sigma_c^{++}$ 
baryons in the initial collision of the antiproton with one of the protons 
of the target nucleus. The medium effects on the exchanged baryons are 
included by incorporating in the corresponding propagators, the effective 
charm baryon masses calculated within a quark-meson coupling (QMC) model. 
The wave functions of the bound proton have been determined within the QMC 
model as well as in a phenomenological model where they are obtained by 
solving the Dirac equation with appropriate scalar and vector potentials. 
The initial- and final-state distortion effects have been approximated by 
using an eikonal approximation-based procedure. Detailed numerical results 
are presented for total and double differential cross sections for the 
${\bar D}^0 D^0$ and $D^-D^+$ production reactions on $^{16}$O and $^{90}$Zr 
targets. It is noted that at ${\bar p}$ beam momenta of interest to the 
${\bar P}ANDA$ experiment, medium effects lead to noticeable enhancements 
in the charm meson production cross sections. 
\end{abstract}
\maketitle

\section{Introduction}

Several interesting and intriguing questions in hadron physics can be elucidated
by experiments involving medium-energy antiproton (${\bar p}$) beams on 
fixed-targets. The future ${\bar P}ANDA$ ("antiproton annihilation at Darmstadt") 
experiment at the under-construction antiproton and ion research facility (FAIR) 
in Darmstadt, Germany, will perform such studies at the beam momenta $\leq$ 
15 GeV/c. The physics program of ${\bar P}ANDA$ experiment~\cite{pan09} includes 
the study of bound states of quantum chromodynamics (QCD) up to the region of charm 
quarks. This will mainly concentrate on experiments on charmonium production, 
open charm spectroscopy, the search for charmed hybrids decaying to ${\bar D} D$, 
the rare decays and the charge-conjugation-parity ($CP$) violation in the 
$D$-meson sector.  For accurate detection of the charmonium states above the 
${\bar D} D$ threshold, reliable estimations are required for the production 
rates of ${\bar D}^0 D^0$ and $D^- D^+$ meson pairs (to be together referred to 
as the ${\bar D} D$ mesons) in ${\bar p}$-induced reactions on proton as well 
as heavier nuclear targets at the appropriate energies. The $\bar{P}ANDA$ experiment 
intends to carry out this task~\cite{har07}.

In a recent publication~\cite{shy16}, calculations have been presented for the cross 
sections of the $\bar p + p \to {\bar D} + D$ reactions in the beam momentum range 
of threshold to 20 GeV/c within a single-channel effective Lagrangian model (ELM) 
(see, {\it e.g.}, Refs. \cite{shy99,shy02,shy11}). In this approach, the dynamics of 
the production process is described by the $t$-channel $\Lambda_c^+$, $\Sigma_c^+$, 
and $\Sigma_c^{++}$ baryon exchange diagrams. The initial- and final-state 
interactions have been accounted for by an eikonal type of phenomenological model. It 
has been found that at beam momenta beyond the threshold region, the total cross 
sections of the ${\bar p} + p \to {\bar D}^0 + D^0$ reaction are dominated by the 
contributions of the $\Lambda_c^+$ baryon exchange. These cross sections peak around 
$p_{{\bar p}}^{lab}$ of 9 GeV/c (with magnitudes close to 1 $\mu$b). At 
$p_{{\bar p}}^{lab}$ around 15 GeV/c, which is of interest to the ${\bar P}ANDA$ 
experiment, the total cross section of this reaction as obtained in Ref.~\cite{shy16},
is  at least 5 times larger than its maximum value predicted in other studies
\cite{kai94,kho12,tit08,gor13,gor09,kro89,hai14}. The large ${\bar D}^0 D^0$ 
production cross section raises the hope of studying the charm mixing, and searching 
for possible new physics contributions via clean signatures of charm $CP$ violation
\cite{kap10}.

On the other hand, in the ELM the $\bar p + p \to  D^- + D^+$ reaction amplitudes 
involve only the $\Sigma_c^{++}$ baryon exchange contribution. They have been 
found to be strongly suppressed compared to those of the ${\bar D}^0 D^0$ 
production. This is attributed to the much smaller coupling constant of the 
$\Sigma_c^{++}$-exchange vertex in comparison to that of the $\Lambda_c^+$-exchange. 
However, in the coupled-channels meson exchange model of Ref.~\cite{hai14}, the 
initial state inelastic interactions could enhance the $D^-D^+$ production 
cross sections significantly.

The studies on the ${\bar D}D$ production in the ${\bar p}$-nucleus collisions are  
expected to explore the properties of the charm hadrons in nuclear medium and provide 
information about their interactions in the nuclear environment (see, e.g, a recent 
review~\cite{tol15}). The ${\bar P}ANDA$ experiment, with the capability of its 
detectors and the energy range of the storage ring, will be able to perform 
measurements for the cross sections of such reactions. The threshold momentum for 
${\bar D}D$ production in ${\bar p}$ induced reaction on proton is 6.40 GeV/c. This 
will be lowered in the ${\bar p}$ reaction on nuclei due to the Fermi motion effects. 
Over the last years some efforts have been made to study theoretically the charm 
production in ${\bar p}$-nucleus reactions within a variety of models (see, e.g., Refs.
\cite{sib99,ger05,lar13}). The latter two studies concentrate on the calculations 
of the production of charmonium states $J/\Psi$ and $\Psi^\prime$. In Ref.~\cite{sib99}, 
${\bar D}D$ meson production in ${\bar p}$-nucleus reaction was investigated within a 
cascade model.

The aim of this paper is to study the ${\bar D}^0 D^0$ and $D^- D^+$ meson-pair 
productions in ${\bar p}$ induced reactions on nuclei within an effective Lagrangian 
model. The basic production mechanism considered in our work is depicted in Fig.~1, 
where the ${\bar p} + A \to {\bar D} + D + B(= A-p)$ reactions proceed via $t$-channel 
$\Lambda_c^+$, $\Sigma_c^+$ and $\Sigma_c^{++}$ baryon exchange diagrams. The 
exchanges of both $\Lambda_c^+$ and $\Sigma_c^+$ baryons contribute to the amplitude 
of the ${\bar p} + A \to {\bar D}^0 + D^0 + B (= A-p)$ reaction. However, the  ${\bar p} 
+ p \to D^- + D^+ + B$ process is mediated only by the exchange of the $\Sigma_c^{++}$ 
baryon.  The $t$-channel part of our model is similar to that of the ${\bar p} + p \to 
{\bar D} + D$ reaction studied in Ref.~\cite{shy16}. It should be mentioned that the 
$s$-channel excitation, propagation and decay into the ${\bar D}D$ channels of the 
$\Psi(3770)$ resonance can also contribute to these reactions. In Ref.~\cite{shy16} it 
was shown that the contributions of the $\Psi(3770)$ resonance to the total cross 
sections of the  ${\bar p} + p \to {\bar D}^0 + D^0$ reaction are insignificant at 
beam momenta away from the threshold region. Since our interest in this work is to 
estimate cross section at beam momenta of interest to the ${\bar P}ANDA$ experiment, 
we have not included such diagrams into our calculations. 

In the next section we present our formalism where details of the ELM are presented 
and input required for making calculations within this model are discussed. The results 
and discussions of our work are given in Sec. III. Finally, the summary and conclusions 
of this study are presented in Sec. IV 

\begin{figure}[t]
\centering
\includegraphics[width=.50\textwidth]{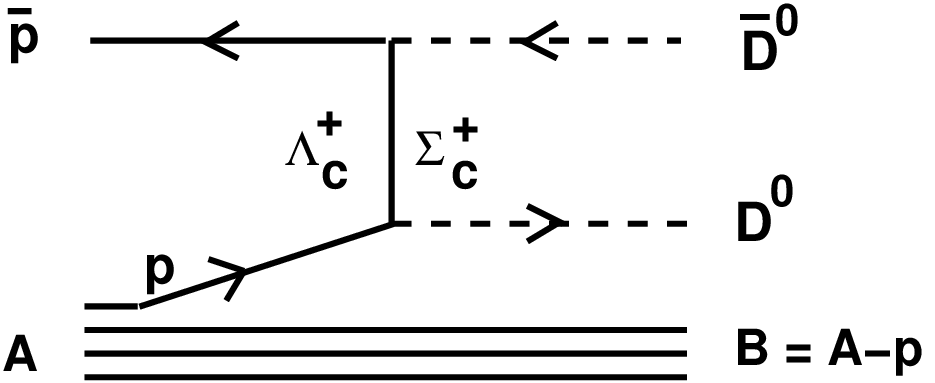}
\caption{
Graphical representation of the model used to describe the ${\bar p} + A 
\to {\bar D}^0 + D^0 + B(= A-p)$ reaction via $t$-channel exchange of charmed 
baryons $\Lambda_c^+$ and $\Sigma_c^+$. Similar representation applies also 
to the  ${\bar p} + A \to D^- + D^+ + (A-p)$ reaction, but the intermediate 
line,  in this case represents the exchange of $\Sigma_c^{++}$ baryon.
Arrows indicate the relative directions of momenta.  
}
\label{fig:Fig1}
\end{figure}

\section{Formalism}

We have followed the procedure and notations of Ref.~\cite{bjo64} in deriving the
formulas for the invariant cross section of the $\bar p + A \rightarrow {\bar D} + D 
+ B$ reaction, which can be written as (see, {\it e.g.}, Refs.~\cite{shy99,shy13}),
\begin{eqnarray}
d\sigma & = & \frac{m_{\bar p} m_A m_B}{\sqrt{[(p_{\bar p} p_{A})^2-
                    m_{\bar p}^2m_A^2]}}
                     \frac{1}{4(2\pi)^5}\delta^4(P_f-P_i)|A_{fi}|^2 
                     \nonumber \\
                 & \times &  \frac{d^3p_{\bar D}}{E_{\bar D}} 
                    \frac{d^3p_{D}}{E_{D}} \frac{d^3p_B}{E_B}, \label{eq.1}
\end{eqnarray}
where $A_{fi}$ represents the total amplitude, $P_i$ and $P_f$ the sum of all the
momenta in the initial and final states, respectively.  $m_{\bar p}$, $m_A$ and 
$m_B$ are the masses of the antiproton, and nuclei A and B, respectively. $p_{\bar p}$
and  $p_{A}$ are the momenta of the antiproton and the target nucleus, respectively. 
The cross sections in the laboratory or CM systems can be written from this equation by 
imposing the relevant conditions. Summations over final spin states and average over 
initial spin states are implied in $|A_{fi}|^2$.

To evaluate amplitudes for the processes represented in Fig.~1, we have used the
effective Lagrangians at the charm baryon-meson-nucleon vertices, which are taken 
from Refs.~\cite{shy14,hob14,hai11a,gro90,wie10}. For the vertices involved in the 
$t$-channel diagrams we have 
\begin{eqnarray}
{\cal L}_{N{C_B}D} & = & ig_{N{C_B}D}{\bar \psi}_N i\gamma^5 \phi_{D}\psi_{C_B} + H.c.,
\label{eq.2}
\end{eqnarray}
\noindent
where ${\psi}_N$ and $\psi_{C_B}$ are the nucleon (antinucleon) and charmed 
baryon ($C_B$) fields, respectively, and $\phi_{D}$ is the $D$-meson field. 
$g_{N{C_B}D}$ in Eq.~(\ref{eq.2}), represents the vertex coupling constant.
For calculating the amplitude of the processes represented in Fig.~1, we require 
the in-medium propagators for the intermediate baryons $C_B$ ($\Lambda_c^+$, 
$\Sigma_c^+$, and $\Sigma_c^{++}$). We write these propagators as, 

\begin{eqnarray}
\mathcal{D}_{C_B}(q_{C_B}) = \frac{\im (\gamma_\mu q_{C_B}^\mu + m_{C_B}^*)}
{q_{C_B}^2 - (m_{C_B}^* - \im\Gamma_{C_B}/2)^2},\label{Eq.3}
\end{eqnarray}
where we have introduced the effective mass of the charmed baryon, $m^*_{C_B}$, 
to take into account the medium effects on the propagation of the charmed baryon 
in the nuclear medium. In Eq.~(\ref{Eq.3}) $q_{C_B}$ and $\Gamma_{C_B}$ are the 
four-momentum and width of the exchanged charmed baryon, respectively.  We have 
calculated $m^*_{C_B}$ within the quark-meson coupling (QMC) model~\cite{gui88}, 
employing the QMC-I version of the model \cite{gui96,sai96}. The details of this 
calculation are given in Sec. II.A. In the following, the free-space (vacuum) 
mass of the exchanged charmed baryon will be represented by $m_{C_B}$. For 
$\Lambda_c^+$ and $\Sigma_c^+$ charmed baryons, the values of $m_{C_B}$ are 
taken to be 2285 and 2452 MeV, respectively. The values of $m^*_{C_B}$ and 
$m_{C_B}$ for $\Sigma_c^{++}$ have been taken to be the same as those of 
$\Sigma_c^+$. 

In Eq.~(\ref{Eq.3}), we have taken the latest Particle Data Group estimates
~\cite{oli14} for the width  $\Gamma_{C_B}$.  It should, however, be noted that the 
medium effects can also modify the widths of the exchange baryons (see, {\it e.g.}, 
Ref.~\cite{liu10}). Nevertheless, because making predictions for the modification 
in the widths of the charmed baryons is at present out of the scope of the QMC model,
we continue to use on-shell widths for the exchanged baryons, which are very small 
in any case.
 
The amplitude of the process depicted in Fig.~1, is given by
\begin{eqnarray}
A_{fi} & = & i \frac{g_{N{C_B}D}^2}{q_{C_B}^2-(m^*_{C_B}-i\Gamma_{C_B}/2)^2}\,\,
{\bar \psi}_{\bar p}
(k_{\bar p}) \,\gamma^5 \, (\gamma_\mu q_{C_B}^\mu + m^*_{C_B})\, \gamma^5\, 
\psi_A(k_p), \label{eq.4}
\end{eqnarray}
\noindent
where $\psi_A(k_p)$ is the spinor for the bound proton in the initial channel. 
It is a four component Dirac spinor, which is the solution of the Dirac equation 
for a bound state problem in the presence of external scalar and vector potential
fields. This is written as (see, e.g., Ref.~\cite{shy06})
\begin{eqnarray}
\psi(k_p) & = & \delta(k_{p0}-E_p)\begin{pmatrix}
                    {f(K_p) {\cal Y}_{\ell 1/2 j}^{m_j} (\hat {k}_p)}\\
                    {-ig(K_p){\cal Y}_{\ell^\prime 1/2 j}^{m_j}
                     (\hat {k}_p)} 
                    \end{pmatrix}. \label{eq.5}
\end{eqnarray}

In our notation $k_p$ represents a four momentum, and $\vec{k_p}$ a three 
momentum. The magnitude of $\vec {k_p}$ is represented by $K_p$, and its 
directions by $\hat {k}_p$. $k_{p0}$ represents the timelike component of 
momentum $k_p$. In Eq.~(\ref{eq.5}), $f(K_p)$ and $g(K_p)$ are the radial parts 
of the upper and lower components of the spinor $\psi(k_p)$, and 
${\cal Y}_{\ell 1/2 j}^{m_j}$ represent the coupled spherical harmonics. The 
latter is given by
\begin{eqnarray}
{\cal Y}_{\ell 1/2 j}^{m_j} & = & <\ell m_\ell 1/2 \mu | j m_j>
                         Y_{\ell m_\ell}(\hat {k}_p) \chi_{\mu},\label{eq.6}
\end{eqnarray}
where $Y_{\ell m_\ell}$ represents the spherical harmonics, and $\chi_{\mu}$ the 
spin-space wave function of a spin-$\frac{1}{2}$ particle. In Eq.~(\ref{eq.5})
$\ell^\prime = 2j - \ell$ with $\ell$ and $j$ being the orbital and total
angular momenta, respectively. 

The coupling constants $g_{NC_BD}$ are adopted from Refs. ~\cite{hai11a,hob14}, 
as $g_{N \Lambda_c^+ D}$ = 13.50, $g_{N \Sigma_c^+ D}$ = 2.69 and 
$g_{N \Sigma_c^{++} D}$ = 2.69.  From these values it is expected that 
$\Lambda_c^+$ will dominate the $t$-channel production amplitudes.
 
The off-shell behavior of the vertices is regulated by a monopole form factor
(see, e.g., Refs.~\cite{shy99,shy02})
\begin{eqnarray}
F_i(q_{{C_B}_i}) & = & \frac{\lambda_i^2-m^{*2}_{{C_B}_i}}{\lambda_i^2 - 
                        q^2_{{C_B}_i}},
\label{Eq.7}
\end{eqnarray}
where index {\it i} represents the {\it i}th exchanged baryon. $\lambda_i$ is 
the corresponding cutoff parameter, which governs the range of suppression of 
the contributions of high momenta carried out via the form factor. We chose a 
value of 3.0 GeV for $\lambda_i$ at all the vertices. The same $\lambda_i$ was 
also used in the monopole form factor employed in the studies of the 
${\bar \Lambda}_c^- \Lambda_c^+$ and ${\bar D} D$ production in the ${\bar p}p$ 
collisions in Refs.~\cite{shy14} and \cite{shy16}, respectively, within a similar 
type of the effective Lagrangian model. Since our calculations are carried out 
in momentum space, they include all the nonlocalities in the production 
amplitudes that arise from the resonance propagators.

We have used plane waves to describe the motions of antiproton and $\bar D$ meson 
in the entrance and outgoing channels, respectively. However, initial and final 
state interactions are approximately accounted for within an eikonal approximation 
based procedure (see section 2.B). 
  
\subsection{Effective charmed baryon mass in nuclear matter within the quark-meson 
coupling model}

A relativistic effective Lagrangian density in QMC-I model for hypernuclei
in mean field approximation, which is used for studying the in-medium modifications of
charmed baryons and production of charmed mesons (in a nucleus),
is given by~\cite{sai07,tsu97,tsu98,tsu03,tsu03a,tsu03b,tsu04}:
\begin{eqnarray}
{\cal L}_{QMC} &=& {\cal L}^N_{QMC} + {\cal L}^Y_{QMC},
\label{eq:LagYQMC} \\
{\cal L}^N_{QMC} &\equiv&  \overline{\psi}_N(\vec{r})
\left[ i \gamma \cdot \partial
- M_N^*(\sigma) - (\, g_\omega \omega(\vec{r})
+ g_\rho \frac{\tau^N_3}{2} b(\vec{r})
+ \frac{e}{2} (1+\tau^N_3) A(\vec{r}) \,) \gamma_0
\right] \psi_N(\vec{r}) \quad \nn \\
  & & - \frac{1}{2}[ (\nabla \sigma(\vec{r}))^2 +
m_{\sigma}^2 \sigma(\vec{r})^2 ]
+ \frac{1}{2}[ (\nabla \omega(\vec{r}))^2 + m_{\omega}^2
\omega(\vec{r})^2 ] \nn \\
 & & + \frac{1}{2}[ (\nabla b(\vec{r}))^2 + m_{\rho}^2 b(\vec{r})^2 ]
+ \frac{1}{2} (\nabla A(\vec{r}))^2, \label{eq:LagN} \\
{\cal L}^Y_{QMC} &\equiv&
\overline{\psi}_Y(\vec{r})
\left[ i \gamma \cdot \partial
- M_Y^*(\sigma)
- (\, g^Y_\omega \omega(\vec{r})
+ g^Y_\rho I^Y_3 b(\vec{r})
+ e Q_Y A(\vec{r}) \,) \gamma_0
\right] \psi_Y(\vec{r}), 
\nn\\
& & (Y = \Lambda,\Sigma^{0,\pm},\Xi^{0,+},
\Lambda^+_c,\Sigma_c^{0,+,++},\Xi_c^{0,+},\Lambda_b),
\label{eq:LagY}
\end{eqnarray}
where $\psi_N(\vec{r})$ and $\psi_Y(\vec{r})$ are the nucleon and the 
hyperon (strange, charm or bottom baryon) fields, respectively.

In an approximation where the $\sigma$, $\omega$ and $\rho$ fields couple
only to the $u$ and $d$ light quarks, the coupling constants for the 
hyperon $Y$, are obtained as $g^Y_\omega = (n_q/3) g_\omega$, and
$g^Y_\rho \equiv g_\rho = g_\rho^q$, with $n_q$ being the total number of
valence light quarks in the hyperon $Y$. $I^Y_3$ and $Q_Y$ are the third 
component of the hyperon isospin operator and its electric charge in units 
of the proton charge, $|e|$, respectively. The field-dependent 
$\sigma$-$N$ and $\sigma$-$Y$ coupling strengths, $g_\sigma(\sigma) \equiv 
g^N_\sigma(\sigma)$ and  $g^Y_\sigma(\sigma)$, appearing in 
Eqs.~(\ref{eq:LagN}) and~(\ref{eq:LagY}), are defined by
%
\begin{eqnarray}
M_N^*(\sigma) &\equiv& M_N - g_\sigma(\sigma)
\sigma(\vec{r}) ,  \\
M_Y^*(\sigma) &\equiv& M_Y - g^Y_\sigma(\sigma)
\sigma(\vec{r}) , \label{effective_mass}
\end{eqnarray}
%
where $M_N$ ($M_Y$) is the free nucleon (hyperon) mass. Note that the 
dependence of these coupling strengths on the applied scalar field must 
be calculated self-consistently within the quark model
\cite{tsu97,tsu98,gui88,sai94,gui96,sai96a}. Hence, unlike quantum 
hadrodynamics (QHD) model~\cite{QHD}, even though $g^Y_\sigma(\sigma) / 
g_\sigma(\sigma)$ may be 2/3 or 1/3 depending on the number of light 
quarks in the hyperon in free space, $\sigma = 0$ (even this is true 
only when their bag radii in free space are exactly the same), this 
will not necessarily be the case in a nuclear medium.

In the following, we consider the limit of infinitely large, uniform 
(symmetric) nuclear matter, where all scalar and vector fields become 
constant. In this limit, we can treat any single hadron (denoted by 
$h$) embedded in the nuclear medium in the same way as we treated 
a hyperon. One simply need to replace ${\cal L}^Y_{QMC}$ in 
Eq.~(\ref{eq:LagY}) by the corresponding Lagrangian density for the 
hadron $h$.

The Dirac equations for the quarks and antiquarks
($q=u$ or $d$, and $Q=s,c$ or $b$, hereafter)
in the bag of hadron $h$ in nuclear matter at the position
$x=(t,\vec{r})$ are given by~\cite{tsu98a,tsu99}:
\begin{eqnarray}
\left[ i \gamma \cdot \partial_x -
(m_q - V^q_\sigma)
\mp \gamma^0
\left( V^q_\omega +
\frac{1}{2} V^q_\rho
\right) \right]
\left( \begin{array}{c} \psi_u(x)  \\
\psi_{\bar{u}}(x) \\ \end{array} \right) &=& 0,
\label{diracu}\\
\left[ i \gamma \cdot \partial_x -
(m_q - V^q_\sigma)
\mp \gamma^0
\left( V^q_\omega -
\frac{1}{2} V^q_\rho
\right) \right]
\left( \begin{array}{c} \psi_d(x)  \\
\psi_{\bar{d}}(x) \\ \end{array} \right) &=& 0,
\label{diracd}\\
\left[ i \gamma \cdot \partial_x - m_{Q} \right]
\psi_{Q} (x)\,\, ({\rm or}\,\, \psi_{\Qbar}(x)) &=& 0,  
\qquad (|\vec{r}|\le {\rm bag~radius}), 
\label{diracQ}
\end{eqnarray}
where we neglect the Coulomb force, and assume SU(2) symmetry for
the light quarks ($q=u=d$). The constant mean-field potentials in 
nuclear matter are defined by, $V^q_\omega \equiv g^q_\omega 
\omega$ and $V^q_\rho \equiv g^q_\rho b$, with $g^q_\sigma$, 
$g^q_\omega$, and $g^q_\rho$ the corresponding quark-meson coupling 
constants. Note that $V^q_\rho = 0$ in symmetric nuclear matter, 
although this is not true in a nucleus where the Coulomb force 
induces the proton and neutron distribution asymmetry even in a 
nucleus with the same numbers of protons and neutrons to give, 
$V^q_\rho \propto (\rho_p - \rho_n) \ne 0$ at a given position 
in a nucleus.

The normalized, static solution for the ground state quarks or 
antiquarks with flavor $f$ in the hadron $h$, may be written as,
$\psi_f (x) = N_f e^{- i \epsilon_f t / R_h^*} \psi_f (\vec{r})$,
where $N_f$ and $\psi_f(\vec{r})$ are the normalization factor and 
the corresponding spin and spatial part of the wave function,
respectively. The bag radius in medium for a hadron $h$ ($R_h^*$)
is determined through the stability condition for the mass of the 
hadron against the variation of the bag radius
\cite{gui88,sai94,gui96}. The eigenenergies in units of $1/R_h^*$ 
are given by,
\begin{eqnarray}
\left( \begin{array}{c}
\epsilon_u \\
\epsilon_{\bar{u}}
\end{array} \right)
= \Omega_q^* \pm R_h^* \left(
V^q_\omega
+ \frac{1}{2} V^q_\rho \right),\,\,
\left( \begin{array}{c} \epsilon_d \\
\epsilon_{\bar{d}}
\end{array} \right)
= \Omega_q^* \pm R_h^* \left(
V^q_\omega
- \frac{1}{2} V^q_\rho \right),\,\,
\epsilon_{Q}
= \epsilon_{\Qbar} =
\Omega_{Q}.
\label{energy}
\end{eqnarray}

The hadron masses in a nuclear medium $m^*_h$ (free mass $m_h$),
are calculated by

\begin{eqnarray}
m_h^* &=& \sum_{j=q,\bar{q},Q,\Qbar}
\frac{ n_j\Omega_j^* - z_h}{R_h^*}
+ \frac{4}{3}\pi R_h^{* 3} B,\quad
\left. \frac{\partial m_h^*}
{\partial R_h}\right|_{R_h = R_h^*} = 0,
\label{hmass}
\end{eqnarray}
%
\begin{table}
\begin{center}
\begin{minipage}[t]{14.5cm}
\caption{Current quark masses (input), coupling constants
and the bag constant.}
\label{coupcc}
\end{minipage}
\begin{tabular}[t]{r|r||r|r}
\hline
$m_{u,d}$ &5    MeV &$g^q_\sigma$ &5.69\\
$m_s$     &250  MeV &$g^q_\omega$ &2.72\\
$m_c$     &1300 MeV &$g^q_\rho$   &9.33\\
$m_b$     &4200 MeV &$B^{1/4}$    &170 MeV\\
\end{tabular}
\end{center}
\end{table}
\begin{table}
\begin{center}
\begin{minipage}[t]{14.5cm}
\caption{The bag parameters,
various hadron masses and the bag radii in free space
[at normal nuclear matter density, $\rho_0=0.15$ fm$^{-3}$]
$z_h, R_h$ and $M_h$ [$M_h^*$ and $R_h^*$].
$M_h$ and $R_N = 0.8$ fm in free space are inputs.
}
\label{bagparambc}
\end{minipage}
\begin{tabular}{c|ccc|cc}
\hline
h &$z_h$ &$M_h$ (MeV) &$R_h$ (fm) &$M_h^*$ (MeV) &$R_h^*$ (fm)\\
\hline
$N$           &3.295 &939.0  &0.800 &754.5  &0.786\\
$\Lambda_c$   &1.766 &2284.9 &0.846 &2162.5 &0.843\\
$\Sigma_c$    &1.033 &2452.0 &0.885 &2330.2 &0.882\\
\end{tabular}
\end{center}
\end{table}
where $\Omega_q^*=\Omega_{\bar{q}}^* =[x_q^2 + (R_h^* m_q^*)^2]^{1/2}$, 
with $m_q^*=m_q{-}g^q_\sigma \sigma$, $\Omega_Q^*=\Omega_{\Qbar}^*=
[x_Q^2 + (R_h^* m_Q)^2]^{1/2}$, and $x_{q,Q}$ being the lowest bag 
eigenfrequencies. $n_q (n_{\qbar})$ and $n_Q (n_{\Qbar})$ are the quark 
(antiquark) numbers for the quark flavors $q$ and $Q$, respectively. The 
MIT bag quantities, $z_h$, $B$, $x_{q,Q}$, and $m_{q,Q}$ are the 
parameters for the sum of the c.m. and gluon fluctuation effects, bag 
constant, lowest eigenvalues for the quarks $q$ or $Q$, respectively, 
and the corresponding current quark masses. $z_N$ and $B$ ($z_h$) are 
fixed by fitting the nucleon (the hadron) mass in free space. For the 
current quark masses we use $(m_{u,d},m_s,m_c,m_b) = (5,250,1300,4200)$ 
MeV, where the values for $m_c$ and $m_b$ are the averaged values from 
Refs.~\cite{PDG96} and~\cite{PDG00}, respectively, and these values were 
used in Refs.~\cite{tsu03,tsu03a,tsu03b,tsu04}. Since, the effects of the 
bare quark mass values used are very small on the results, we use the 
same values as used in the past so that we can compare and discuss the 
results with those obtained previously~\cite{tsu03,tsu03a,tsu03b,tsu04}.
This also applies for the baryon mass values used. The bag constant 
calculated for the present study is $B = (170$ MeV$)^4$.
The quark-meson coupling constants, which are determined so as
to reproduce the saturation properties of symmetric nuclear matter (the 
binding energy per nucleon of 15.7 MeV at $\rho_0 = 0.15$ fm$^{-3}$),
are ($g^q_\sigma, g^q_\omega, g^q_\rho$) = ($5.69, 2.72, 9.33$), where 
$g_\sigma \equiv g^N_\sigma \equiv 3 g^q_\sigma S_N(0) = 3 \times 5.69 
\times 0.483 = 8.23$~\cite{sai96a}. These are summarized in Table~
\ref{coupcc}. The parameters $z_h$, and the bag radii $R_h$ for relevant 
baryons in free space, and some quantities calculated at normal nuclear 
mater density $\rho_0 = 0.15$ fm$^{-3}$ are listed in 
Table~\ref{bagparambc}, together with the free space masses 
(inputs)~\cite{PDG96,PDG00,PDG98,PDG02}.
\begin{figure*}
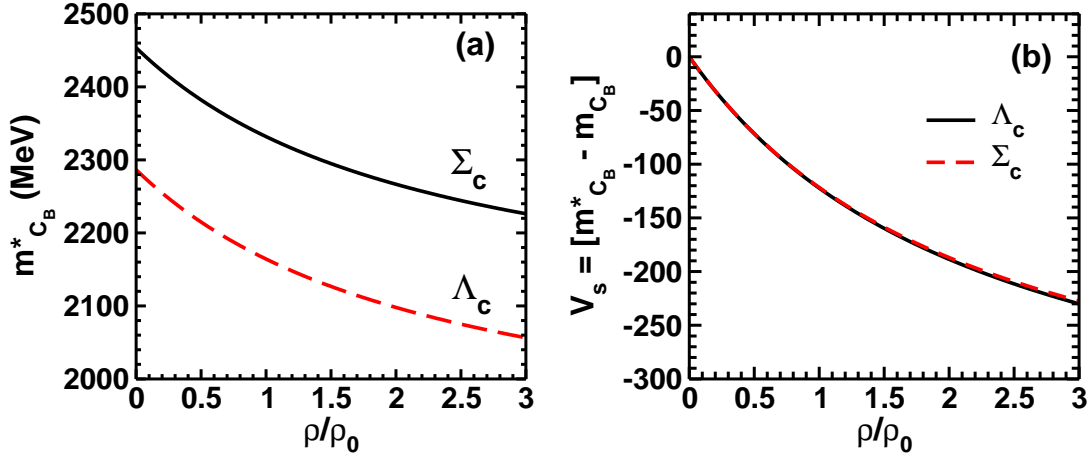

\begin{tabular}{cc}
\includegraphics[scale=0.50]{medium_mass-Fig2a.eps} & \hspace{0.20cm}
\includegraphics[scale=0.50]{scalar_pot-Fig2b.eps}
\end{tabular}
\caption{[Color online]
(a) Effective masses of the charmed baryons $\Lambda_c^+$ and $\Sigma_c^+$ in  
symmetric nuclear matter as a function of matter density, as calculated within 
the QMC-I model. (b) Scalar potential for $\Lambda_c^+$ and $\Sigma_c^+$ 
charmed baryons as a function of a matter density in symmetric nuclear matter.
}
\label{Fig2}
\end{figure*}

In Fig.~\ref{Fig2} (a), we show  the effective masses of the charmed baryons $C_B$,   
and in Fig.~\ref{Fig2} (b) the scalar potentials in symmetric nuclear matter as a 
function of nuclear density. We note that $m^*_{C_B}$ $\le$ $m_{C_B}$ at finite 
density as usually expected. The attractive mass-shift that reflects the reduced 
light-quark condensates at finite density, have been predicted for vector mesons 
as well~\cite{bro91,hat92}, although the issues of in-medium widths related with 
the collision broadening must be studied carefully in the experimental situations.
Our calculations predict the similar observation for the charm baryon sector 
as those for the strange sector and the nucleons. We further note that in-medium 
mass shift of charmed baryons leads to an attractive scalar potential of about 
120.00 MeV at normal nuclear matter density $\rho = \rho_0$.

\subsection{Initial- and final-state interactions}

From the studies of the ${\bar \Lambda}_c^- \Lambda_c^+$ and ${\bar D } D$ 
production~\cite{hai10,shy14,shy16} in the ${\bar p} p$ collisions, it was found 
that the magnitudes of the cross sections depend very sensitively on the 
initial-state distortion effects. In fact, the ${\bar p} p$ annihilation channel 
is almost as strong as the elastic scattering channel. Consequently, in 
${\bar p}$-nucleus reactions also these distortion effects are expected to be as 
significant. They can produce both absorptive as well as dispersive effects. 
However, for large incident energies involved in this study, the absorptive 
effects are likely to be most important. We estimate these  within an eikonal 
approximation, as discussed below.
 
Within the eikonal approximation, the attenuation factor for a particle 
traveling through the nuclear medium can be written as (see, e.g. Ref.
\cite{shy06}) 
\begin{eqnarray}
S(E) & = & \int\, d{\bf b}\,dz\, \frac{\rho({\bf b},z)exp[-k\,\eta_0(E)
         L(b)]}{\int\,d{\bf b}\,dz\,\rho({\bf b},z)},\label{eq.8}
\end{eqnarray}
where $\eta_0$ is the imaginary part of the refractive index of the 
nuclear medium and $\rho(r) [\rho(\sqrt{b^2+z^2})]$ is the nuclear density 
distribution, with $b$ being the impact parameter. In Eq.~(\ref{eq.8}) $L(b)$ 
is the length of the path traveled by the particle in the medium, which is 
given by
\begin{eqnarray}
L(b) & = & \int_0^\infty \frac{\rho(r)}{\rho_0}\,dz. \label{eq.9}
\end{eqnarray}
If the nuclear density is approximated by a Gaussian function, $\rho(r) = 
\rho_0\,exp\,(-r^2/\alpha^2)$, the integration in Eq.~(\ref{eq.8}) can be done 
analytically.  In this case the attenuation factor is given by
\begin{eqnarray}
S(E) & = & \frac{1\,-\,exp[\,-\,\sqrt{\pi}\,\alpha\,k\,\eta_0(E)]}{
         \sqrt{\pi}\alpha \,k\,\eta_0(E)}.\label{eq.10}
\end{eqnarray}
The attenuation due to medium can be calculated if the value of $\eta_0(E)$ 
is known. This can be obtained from the imaginary part of the optical potential 
$W_0$ as
\begin{eqnarray}
\eta_0(E) & = & \frac{1}{\hbar^2}\frac{E}{k^2}W_0\,(E).\label{eq.11}
\end{eqnarray}
One can use the following high-energy relations to relate $W_0$ to the 
${\bar p}p $ total cross section, $\sigma_T$  
\begin{eqnarray}
W_0\,(E) & = & \hbar^2 \frac{k\,\sigma_T \rho_0}{2E}.\label{eq.13}
\end{eqnarray}

In order to determine the total reduction factor for the ${\bar p} + A \to 
{\bar D}^0 + D^0 + B(= A-p)$ reaction, the total attenuation due to both the 
antiproton and $D$-meson distortions has been estimated by replacing the factor 
$k\eta_0$ in Eq.~(\ref{eq.10}) by
\begin{eqnarray}
k\eta_0 &  \rightarrow & k_{\bar p} \eta_0(E_{\bar p}) + k_{\bar D} 
\eta_0(E_{\bar D}) + k_D \eta_0(E_D).\label{eq.14}
\end{eqnarray}

The information about the ${\bar p}$-nucleus imaginary potential is not very 
firm~\cite{gai11,hrt16} particularly at higher beam mometa of interest to the
${\bar P}ANDA$ experiment. In our estimation of the attenuation we have taken 
the average value of the imaginary part of the ${\bar p}$-nucleus optical 
potential to be 125 MeV, which is in agreement  with the values reported in 
Refs.~\cite{zha96,lar16} at higher antiproton momenta. We have made this 
value independent of the beam momentum and the target nucleus. This corresponds 
to a ${\bar p}p$ $\sigma_T$ of about 75 mb at $\bar p$ momentum of 15 GeV/c$^2$, 
with a $\rho_0$ of 0.15 fm$^{-3}$. This value is somewhat larger than those 
reported in Refs.~\cite{hai08} and ~\cite{tec14}.

At the same time, the knowledge about the ${\bar D}$- and $D$-nucleus potentials 
is extremely scarce. We have taken a value of 10 mb for ${\bar D}N$ and ${D}N$ 
$\sigma_T$~\cite{tec14,yam16}. The value of the parameter $\alpha$ is taken to be 
2.73 fm and 4.30 fm for $^{16}$C and $^{90}$Zr targets, respectively~\cite{bar77}. 
\begin{figure*}
\centering
\includegraphics[width=.50\textwidth]{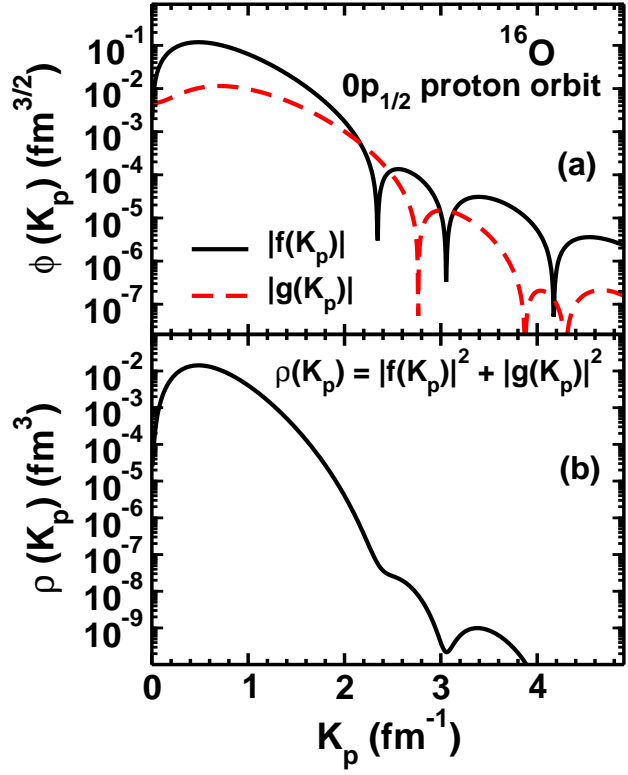}
\caption{(color online)
(a) Momentum space spinors ($\phi(K_p)$) for $0p_{1/2}$ nucleon orbit in $^{16}$O.
$f(K_p)$ and $g(K_p)$ are the upper and lower components of the spinor, respectively.
(b) Momentum distribution ($\rho(K_p)$) for the same state calculated with the wave
function shown in (a).
}
\label{fig:Fig3}
\end{figure*}

\section{Results and Discussions}

The initial bound proton spinors (corresponding to momenta $K_p$) are required 
to perform numerical calculations of the amplitudes in Eq.~(\ref{eq.4}). To 
simplify the nuclear structure problems, we assume the bound proton  states to 
have a pure single particle-hole configuration (with the core remaining 
inert), having quantum numbers of the outermost proton orbit of the target 
nucleus, even though it is straightforward to include also those cases where 
the participating proton occupies other orbits. This corresponds to the 
$0p_{1/2}$ orbit with a binding energy of 12.13 MeV in case of the 
$^{16}$O target, and the $1p_{1/2}$ orbit with a binding energy of 8.35 MeV 
for $^{90}$Zr target.

The spinors in the momentum space are obtained by Fourier transformation of 
the corresponding coordinate space spinors, which are the solutions of the 
Dirac equation with potential fields consisting of an attractive scalar part 
and a repulsive vector part having a Woods-Saxon form. This choice appears 
to be justified as the Dirac Hartree-Fock calculations~\cite{mil72,bro78} 
suggest that these potentials tend to follow the nuclear shape. 
\begin{table}
\begin{center}
\caption[T2] {Searched depths of vector and scalar potentials and the binding
energies of the nucleon bound states.
}
\vspace{1.0cm}
\begin{tabular}{|c|c|c|c|c|c|c|c|} \hline
 State & Binding Energy ($\epsilon$)& $V_s$ & $r_s$ & $a_s$ & $V_v$ & $r_v$ & $a_v$\\
       &(\footnotesize{MeV})&(\footnotesize{MeV})&(\footnotesize{fm})&
  (\footnotesize{fm}) & (\footnotesize{MeV}) & (\footnotesize{fm}) &
(\footnotesize{fm}) \\
\hline
$^{16}$O$(0p_{1/2})$ & 12.13 & -445.56 & 0.983 & 0.606 & 360.91 &
0.983 & 0.606  \\ 
$^{90}$Zr$(1p_{1/2})$ & 8.35 & -418.55 & 0.983 & 0.606 & 339.03 &
0.983 & 0.606  \\ 
\hline
\end{tabular}
\end{center}
\end{table}

In the phenomenological model, the potential fields were obtained by a well-depth
search procedure. In this method, with fixed geometry parameters (radius and 
diffuseness), the depths of the scalar ($V_s$) and ($V_v$) potentials are 
searched to reproduce the binding energies of the respective proton bound states 
with the given choice of quantum numbers. For the target nuclei, $^{16}$O, and 
$^{90}$Zr, the resulting values are shown in Table III. To show the momentum spread 
of the corresponding spinors, we have displayed in Figs.~3 and 4, the spinors  
$|f(K_p)|$ and $|g(K_p)|$ and the momentum distribution $\rho (K_p) = |f(K_p)|^2 
+ |g(K_p)|^2$  as a function of momentum $K_p$ for $^{16}$O and $^{90}$Zr targets, 
respectively. It was shown in Ref.~\cite{shy95} that spinors calculated in this way 
provide a good description of the nucleon momentum distribution for the $p$ shell 
nucleons. We note that in the region of momentum transfer pertinent to charm-meson 
production in ${\bar p}$-nucleus collisions, the lower components of the spinors are 
not negligible as compared to the upper component, which clearly demonstrates that 
a fully relativistic approach is essential for an accurate description of this 
reaction.
\begin{figure}[t]
\centering
\includegraphics[width=.50\textwidth]{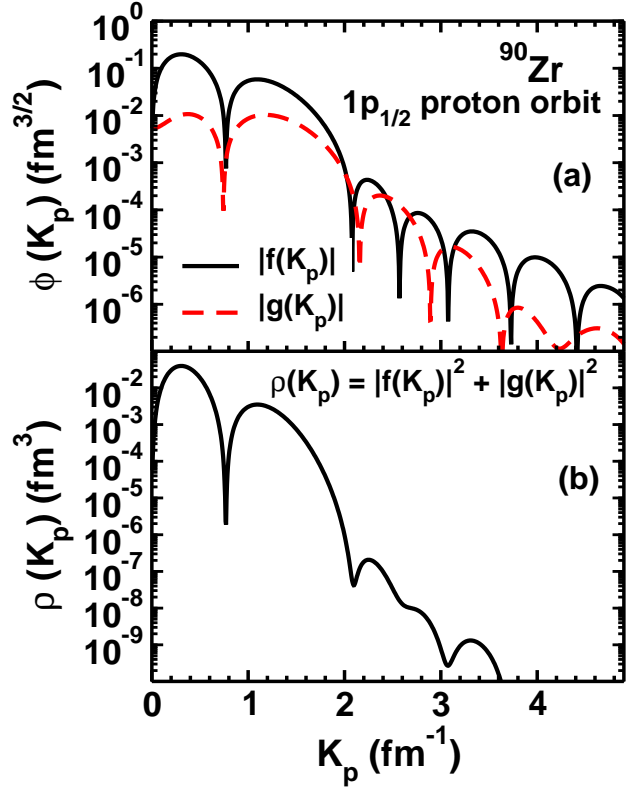}
\caption{(color online)
(a) Momentum space spinors ($\phi(K_p)$) for $1p_{1/2}$ nucleon orbit in $^{90}$Zr.
$f(K_p)$ and $g(K_p)$ are the upper and lower components of the spinor, respectively.
(b) Momentum distribution ($\rho(K_p)$) for the same state calculated with the wave
function shown in (a).
}
\label{fig:Fig4}
\end{figure}

The bound state spinors have also been calculated within the QMC-I model (see, 
{\it e.g.,} Ref.~\cite{shy12}). Even though the binding energies of the bound 
proton orbitals predicted by the QMC model were somewhat different from those 
used in the fitting procedure of the phenomenological model, the spinors 
obtained in two models were almost identical to each other. 

Using the formalism, approximations and input parameters given in Sec. II and 
the bound state spinors described above, we have investigated the  ${\bar D}^0 D^0$ 
and $ D^- D^+$ production in the ${\bar p}$ collision with a light ($A$ = $^{16}$O) 
and a medium mass ($A$ = $^{90}$Zr) nucleus.  We emphasize that parameters like 
coupling constants at the vertices and the shapes of the form factors and the values
of the cutoff parameters involved therein, were the same as those used in the studies 
of the ${\bar D}^0 D^0$ and $D^- D^+$ charmed-meson productions in the ${\bar p} p$ 
collisions at the beam momenta ranging from the corresponding threshold to 20 GeV/c 
in Refs.~\cite{shy16} and \cite{shy14}, respectively. In each case the effects of 
initial- and final-state interactions are included by following the procedure 
described in Sec. II.B.

In Fig.~5, we display the beam momentum dependence of the cross section 
$d\sigma/d\Omega_{{\bar D}^0}$ at $\theta_{{\bar D}^0} = 0^\circ$ for the 
${\bar p}$ + $^{16}$O $\to {\bar D}^0 + D^0 + ^{15}$N reaction . In this figure the 
arrow shows the threshold for the ${\bar D}^0 D^0$ production in ${\bar p} p$ 
collision, which is 6.4 GeV/c. In contrast, the thresholds of this reaction on 
$^{16}$O and $^{90}$Zr targets are 2.89 GeV/c and 2.68 GeV/c, respectively. The shift 
in the threshold of ${\bar p}$-nucleus reactions is mainly due to Fermi motion effects. 

The solid line in Fig.~5 shows the cross section obtained by using in the reaction 
amplitudes the in-medium effective mass ($m_{C_B}^*$) of the exchanged charmed baryon 
(ECB) that are calculated self-consistently within the QMC-I model as discussed in 
section II.A. The dotted line represents the results obtained with the free-space 
(vacuum) mass ($m_{C_B}$) for ECB. It is clear that at the ${\bar p}$ beam momenta 
($p_{\bar p}$) of interest to the ${\bar P}ANDA$ experiment (around 15 GeV/c), the 
effect of nuclear medium is noticeable.  In this region the cross sections calculated 
with the effective mass $m_{C_B}^*$ of ECB are about a factor of 2 larger than 
those obtained with the vacuum mass $m_{C_B}$. 

\begin{figure}[t]
\centering
\includegraphics[width=.50\textwidth]{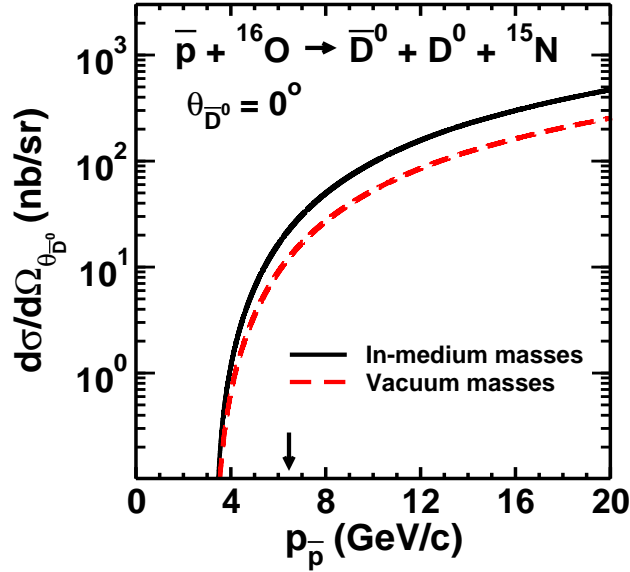}
\caption{(color online)
Differential cross section $d\sigma/d\Omega_{{\bar D}^0}$ at $\theta_{{\bar D}^0} 
= 0^\circ$ as function of antiproton beam momentum for the reaction ${\bar p}$ +
$^{16}$O $\to {\bar D}^0 + D^0 + ^{15}$N. The results obtained with the in-medium 
effective mass ($m_{C_B}^*$) of the exchanged charm-baryons are shown by the full 
line, while those with the free-space (vacuum) mass $m_{C_B}$ are represented by 
the dashed line. In both cases the contributions of $\Lambda_c^+$ and $\Sigma_c^+$ 
(intermediate states) are coherently summed in the reaction amplitude. 
}
\label{fig:Fig5}
\end{figure}

In the amplitudes corresponding to the cross sections shown in Fig. 5, the individual
contributions of the $\Lambda_c^+$ and $\Sigma_c^+$ exchanges have been coherently 
summed. However, we have noted that these cross sections  are almost solely governed 
by the $\Lambda_c^+$-exchange mechanism in the entire range of the antiproton beam 
momentum. The contributions of $\Sigma_c^+$-exchange terms are lower by about 3 orders 
of magnitude. This reflects the trend seen in the case of ${\bar D}^0 D^0$ production 
in ${\bar p} p $ collisions in Ref.~\cite{shy16}.  This can be understood from the 
fact that the coupling constants of the vertices involved in the $\Sigma_c^+$-exchange 
are much smaller than those of the $\Lambda_c^+$-exchange.  

In Fig.~6, we show the cross sections $d\sigma/d\Omega_{D^+}$ at $\theta_{D^+} = 
0^\circ$ for the ${\bar p}$ + $^{16}$O $\to {D^-} + D^+ + ^{15}$N reaction as a 
function of ${\bar p}$ beam momentum. We see that in this case the cross sections are 
strongly suppressed compared to those of Fig.~5. The cross sections of the ${\bar p} 
+ p \to D^- + D^+$ reaction also were found to be similarly suppressed as compared 
to those of the ${\bar p} + p \to {\bar D}^0 + D^0$ reaction in Ref.~\cite{shy16} 
where calculations were performed within a similar ELM model. Same trend was also 
observed in the calculations presented in Refs.~\cite{kai94,tit08,kho12} within models 
that use the idea of Regge trajectory-exchange. In both the cases, this effect can be 
understood from the fact that while the ${\bar D}^0 D^0$ production is dominated by the 
$\Lambda_c^+$-exchange mechanism, the $D^- D^+$ production gets contributions only from 
the $\Sigma_c^{++}$-exchange terms. The coupling constants in the latter case have been 
taken to be equivalent to those of the $\Sigma_c^+$-exchange vertices. Threfore, they 
are much smaller than those of the $\Lambda_c^{+}$-exchange terms. The ratio of the 
absolute magnitudes of the  ${\bar D}^0 D^0$ and $D^- D^+$ production cross sections 
is roughly proportional to $(g_{N\Lambda_c^+D}/g_{N\Sigma_c^{++}D})^4$, which leads 
to a reduction in the $D^-D^+$ production cross section over that of ${\bar D}^0 D^0$ 
by nearly a factor of 650.
\begin{figure}[t]
\centering
\includegraphics[width=.50\textwidth]{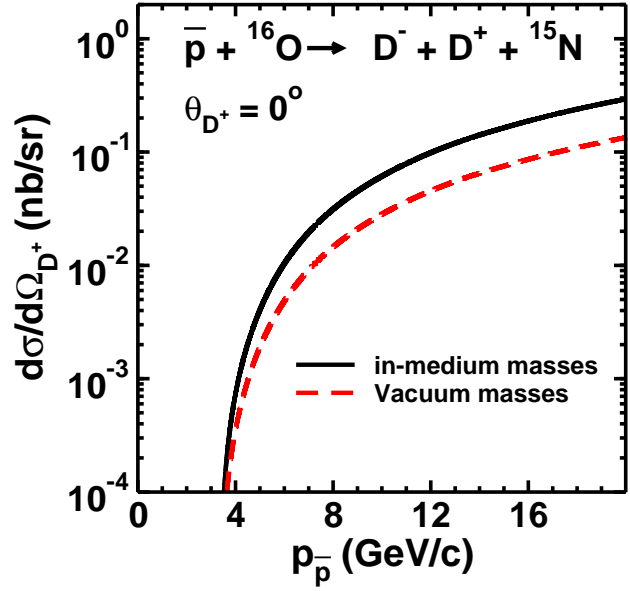}
\caption{(color online)
Differential cross section $d\sigma/d\Omega_{D^+}$ at $\theta_{D^+} = 0^\circ$ as 
function of antiproton beam momentum for the reaction ${\bar p}$ + $^{16}$O $\to 
{D^-} + D^+ + ^{15}$N. The solid and dashed lines have the same meaning as in Fig.~5. 
}
\label{fig:Fig6}
\end{figure}

However, in the coupled-channels meson-exchange model of Ref.~\cite{hai14}, the 
${\bar p} + p \to D^- + D^+$ cross sections are even larger than the 
${\bar p} + p \to {\bar D}^0 + D^0$ ones. This results from their coupled-channels 
treatment of the incident channel, which accounts effectively for two-step 
inelastic processes involving $\Lambda^+_c$ "baryon exchange". Such two-step 
mechanisms are out of the scope of our ELM as well as of the Regge model
\cite{kai94,tit08,kho12} calculations.  Therefore, in studies within these 
models the $D^-D^+$ production is suppressed as compared to the ${\bar D}^0 D^0$ 
production reaction. It should be mentioned here that the cross sections for the 
${\bar p} + p \to D^- + D^+$ reaction from Ref.~\cite{hai14} were used as input in 
Ref.~\cite{yam16} in the calculations of the formation cross sections of the 
$D$-mesic nucleus $[^{11}$B$ - D^-]$ via the reaction ${\bar p} + ^{12}$C $\to 
[^{11}$B$ - D^-] + D^+$ within a Green's function method. The magnitudes of the 
formation cross sections predicted in Ref.~\cite{yam16} will be strongly 
suppressed if our cross sections for the ${\bar p} + p \to D^- + D^+$ reaction are 
used as input. 
\begin{figure}[t]
\centering
\includegraphics[width=.50\textwidth]{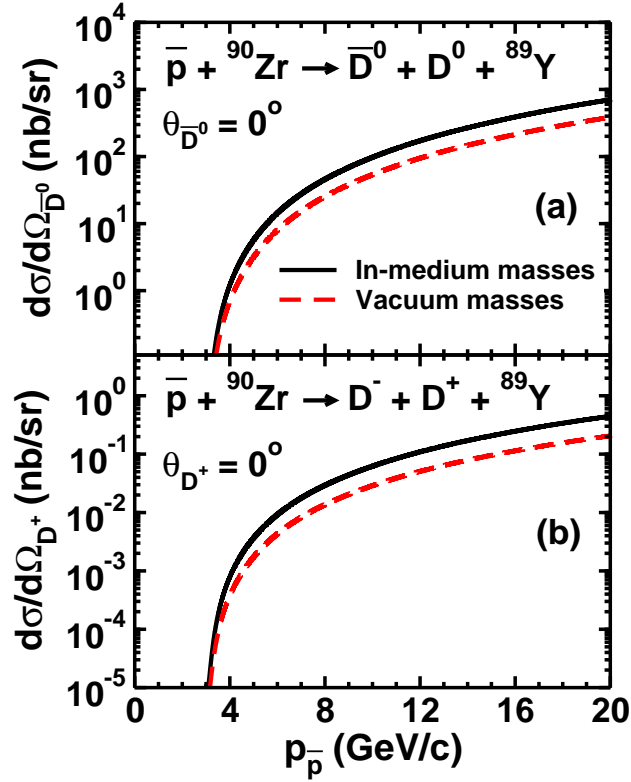}
\caption{(color online)
(a) Differential cross section $d\sigma/d\Omega_{{\bar D}^0}$ at $\theta_{{\bar D}^0} 
= 0^\circ$ as function of antiproton beam momentum for the reaction ${\bar p}$ +
$^{90}$Zr $\to {\bar D}^0 + D^0 + ^{89}$Y. The results obtained with the in-medium
effective mass ($m_{C_B}^*$) of the exchanged charm-baryons are shown by the full 
line while those with the free-space mass $m_{C_B}$ by the dashed line. 
(b) Differential cross section $d\sigma/d\Omega_{D^+}$ at $\theta_{D^+} = 0^\circ$ as 
function of antiproton beam momentum for the reaction ${\bar p}$ + $^{90}$Zr $\to 
{D^-} + D^+ + ^{89}$Y. The solid and dashed lines have the same meaning as in Fig.~7(a). 
}
\label{fig:Fig7}
\end{figure}
  
The solid and dashed lines in Fig.~6 correspond to calculations performed 
with the effective in-medium mass $m^*_{C_B}$ and the vacuum mass $m_{C_B}$, 
respectively. We see that for this reaction too, for $p_{\bar p}$ around 15 
GeV/c, the cross section obtained with the in-medium mass $m^*_{C_B}$ are larger 
than those obtained with $m_{C_B}$ by about a factor of two.

In Fig.~7(a) and 7(b), we display the cross sections $d\sigma/d\Omega_{{\bar D}^0}$ 
at $\theta_{{\bar D}^0} = 0^\circ$ for the ${\bar p}$ + $^{90}$Zr $\to {\bar D}^0 + 
D^0 + ^{89}$Y reaction, and $d\sigma/d\Omega_{D^+}$ at $\theta_{D^+} = 0^\circ$ for
the  ${\bar p}$ + $^{90}$Zr $\to {D^-} + D^+ + ^{89}$Y reaction, respectively, as a 
function of $p_{\bar p}$. The solid and dashed lines have the same meaning as those 
in Figs. 5 and 6. We note that all the features of the cross sections that were 
observed in case of the reactions on the lighter $^{16}$O target are also present in 
those on this medium mass target. There is, however, one difference. While the cross 
sections on the $^{90}$Zr and $^{16}$O targets are approximately similar for 
$p_{\bar p}$ $\leq$ 10 GeV/c, they start differing from each other for $p_{\bar p}$ 
$\ge$ 10 GeV/c.  In this region the cross section on the heavier target becomes 
gradually larger than those on the lighter one.  

\begin{figure}[t]
\centering
\includegraphics[width=.50\textwidth]{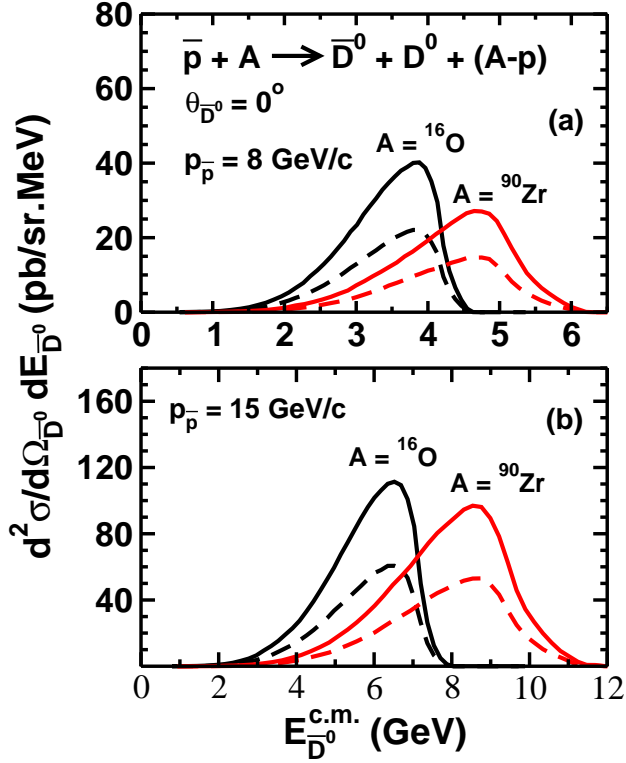}
\caption{(color online)
(a) Double differential cross section $d^2\sigma/d\Omega_{{\bar D}^0}dE_{{\bar D}^0}$
at $\theta_{{\bar D}^0} = 0^\circ$ for the ${\bar p}$ + $A$ $\to$ ${\bar D}^0 + D^0 + 
(A-p)$ reaction where A = $^{16}$O and $^{90}$Zr, as a function of ${\bar D}^0$ energy 
$E_{{\bar D}^0}$ at the ${\bar p}$ beam momentum of 8 GeV/c. The solid and dashed 
lines represent the results obtained with the in-medium effective charmed-baryon mass 
and the free-space charmed baryon mass, respectively. (b) The same as in Fig. 8(a) but 
at the ${\bar p}$ beam momentum of 15 GeV/c.
} 
\label{fig:Fig8}
\end{figure}

In Figs.~8(a) and 8(b),  we present the results for the double differential cross 
section $d^2\sigma/d\Omega_{{\bar D}^0}dE_{{\bar D}^0}$ at $\theta_{{\bar D}^0} = 
0^\circ$ for the ${\bar D}^0 D^0$ production in ${\bar p}$ induced reaction on 
$^{16}$O and $^{90}$Zr targets at $p_{\bar p}$ of 8 GeV/c and 15 GeV/c, respectively, 
as a function of the center-of-mass (c.m.) energy, $E_{{\bar D}^0}^{c.m.}$, of 
${\bar D}^0$ charmed meson. The solid and the dashed lines represent the results 
obtained by using masses $m^*_{C_B}$ and $m_{C_B}$, respectively, for the ECB in 
the calculations.  It is seen that these cross sections are peaked very close to 
the maximum allowed values of $E_{{\bar D}^0}^{c.m.}$ ($E_{{\bar D}^0}^{c.m.,max}$)
corresponding to the given target and the beam momentum. This effect is found for 
reactions on both the targets. As the target mass increases, 
$E_{{\bar D}^0}^{c.m.,max}$ shifts to higher values so does the peak position in the 
corresponding cross section. The widths of the distributions are some what larger 
for the heavier target.

Although the effect of using the in-medium effective mass of the exchanged 
charmed baryon is visible in the entire ${\bar D}^0$ energy spectrum, this is more 
prominent in the region around the peak position, where peak cross sections obtained 
by using in-medium mass $m^*_{C_B}$ in the amplitude are larger by nearly a factor 
of two than those obtained with the free-space mass $m_{C_B}$.  

Therefore, future measurements of the ${\bar D} D$ production in ${\bar p}$ induced 
reactions on nuclei to be performed with the ${\bar P}ANDA$ detector at the FAIR 
facility are expected to provide a handle to probe the in-medium properties of the 
charmed-baryons.

Finally we acknowledge that a major source of uncertainty of our results is  
provided by our treatment of the initial- and final-state interactions. We 
account for these effects within an eikonal-approximation-based phenomenological 
method. Although, the parameters of our method (${\bar p}N$ total cross section and 
the radius parameter of the target nuclei) can be checked from the independent
sources, they should ideally be constrained by fitting to the experimental data.
Because of the lack of any experimental information, it is not yet possible to test
our model thoroughly. Thus the absolute magnitudes of our cross sections may have 
some uncertainties. Furthermore, in the present treatment we considered the 
absorptive distortion effects only that influence the absolute magnitudes of the 
cross sections. In a more rigorous treatment, the inclusion of dispersive effects  
may affect the shapes of the cross sections also.

\section{Summary and conclusions}

In Summary, we have studied the production of charmed-meson pairs ${\bar D}^0 D^0$
and $D^- D^+$ in antiproton induced reactions on $^{16}$O and $^{90}$Zr targets by 
using a single-channel effective Lagrangian model that involves meson-baryon degrees 
of freedom. The dynamics of the production process is described by the t-channel
diagrams involving exchanges of charmed baryons $\Lambda_c^+$, $\Sigma_c^+$, and 
${\Sigma_c^{++}}$ during the collision between the antiproton and a proton bound in 
the target. The initial- and final-state interactions have been taken into account 
by an eikonal type of phenomenological model. The coupling constants at the 
charmed-baryon exchange vertices were taken to be the same as those used in the 
studies of ${\bar p} + p \to  {\bar D} + D$ and  ${\bar p} + p \to{\bar {\Lambda}}_c^-
 + {\Lambda}_c^+$ reactions in Refs.~\cite{shy16} and \cite{shy14}, respectively. 
These coupling constants were deduced in Ref.~\cite{hai11a} from the analysis of the 
$DN$ and ${\bar D}N$ scatterings. The same coupling constants were also used for the 
vertex couplings involved in the $D$-meson-nucleon interactions in the studies 
reported in Ref.~\cite{hob14}. The off-shell corrections at various vertices have 
been accounted for by introducing monopole form factors with the cut-off parameters 
having the same value as those used in our studies reported in Refs.~\cite{shy14} and 
\cite{shy16}. It is a general practice, however, to determine the shape of the form 
factors and the cutoff parameters involve therein by fitting to the experimental 
data. Because, such data are not yet available for the reactions under study in this 
paper, we restricted ourselves to the choice of the form factor and the cutoff 
parameter that were used in our previous study of this reaction on a proton target.
The bound proton spinors have been  obtained by solving the Dirac equation with 
vector and scalar potential fields having Woods-Saxon shapes. Their depths are 
fitted to the binding energy of the respective state.

We find that the differential cross sections for the production of ${\bar D}^0 D^0$ 
charmed-meson pair at the ${\bar D}^0$ angle of $0^\circ$ in ${\bar p}$-induced 
reaction on both $^{16}$O and $^{90}$Zr targets, are dominated by the contributions
of the $\Lambda_c^+$ baryon exchange - the $\Sigma_c^+$-exchange contributions are 
quite small due to relatively smaller coupling constants. The cross sections of the
$D^- D^+$ production that gets contributions solely from the $\Sigma_c^{++}$ baryon 
exchange process, are strongly suppressed due the smaller coupling constants of 
the corresponding vertices.   

The double differential cross sections for the ${\bar p} + A \to {\bar D}^0 + 
D^0$ reaction for observing ${\bar D}^0$ at $0^\circ$ have maxima in the vicinity 
of the kinematically allowed maximum values of ${\bar D}^0$ c.m. energies. This 
feature is independent of the ${\bar p}$ beam momentum and the target mass. The 
widths of the corresponding spectra are, however, target-mass dependent.  

A significant result of our study is that using in-medium effective masses in the 
propagators of the exchanged charmed baryons leads to about a factor 2 increase 
in the cross sections at most forward angle over those obtained with the 
corresponding free-space masses, for antiproton beam momenta around 8-15 GeV/c, 
which are of interest to the ${\bar P}ANDA$ experiment. This result is 
independent of the mass of the target nucleus. This observation suggests that 
in-medium properties of the charmed baryon may be experimentally accessible 
in this experiment.

\section{acknowledgments}
 
This work has been supported by the Science and Engineering Research Board 
(SERB), Department of Science and Technology, Government of India under 
Grant no. SB/S2/HEP-024/2013, and by Funda\c{c}\~{a}o de Amparo \`{a} 
Pesquisa do Estado de S\~{a}o Paulo (FAPESP), Brazil, Grants No.~2016/04191-3, 
and, No.~2015/17234-0, and Conselvo Nacional de Desenvolvimento Cientifico e 
Tecnol\`{o}gico (CNPq), Grants No.~400826/2014-3 and No.~308088/2015-8.

\end{document}